\documentclass[preprint, showpacs,preprintnumbers,amsmath,amssymb,nofootinbib]{revtex4}
\usepackage{epsf}
 \usepackage{graphicx}
\usepackage{float}
\begin{document}
\newcommand{\be}[1]{\begin{equation}\label{#1}}
 \newcommand{\ee}{\end{equation}}
 \newcommand{\bea}{\begin{eqnarray}}
 \newcommand{\eea}{\end{eqnarray}}
 \newcommand{\nn}{\nonumber}

 \def\disp{\displaystyle}
 \def\gsim{ \lower .75ex \hbox{$\sim$} \llap{\raise .27ex \hbox{$>$}} }
 \def\lsim{ \lower .75ex \hbox{$\sim$} \llap{\raise .27ex \hbox{$<$}} }

\title{\Large \bf Detecting the cosmic acceleration with current data}
\author{Rong-Gen Cai~\footnote{Email: cairg@itp.ac.cn}, Zhong-Liang Tuo~\footnote{Email: tuozhl@itp.ac.cn} }

\address{Key Laboratory of Frontiers in Theoretical Physics,
    Institute of Theoretical Physics, Chinese Academy of Sciences,
    P.O. Box 2735, Beijing 100190, China}

\begin{abstract}
The deceleration parameter $q$ as the diagnostic of the cosmological
accelerating expansion is investigated. By expanding the luminosity
distance to the fourth order of redshift and the so-called
$y$-redshift in two redshift bins and fitting the SNIa data
(Union2), the marginalized likelihood distribution of the current
deceleration parameter shows that the cosmic acceleration is still
increasing, but there might be a tendency that the cosmic
acceleration will slow down in the near future. We also fit the
Hubble evolution data together with SNIa data by expanding the
Hubble parameter to the third order, showing that the present
decelerating expansion is excluded within $2\sigma$ error. Further
exploration on this problem is also approached in a
non-parametrization method by directly reconstructing the
deceleration parameter from the distance modulus of SNIa, which
depends neither on the validity of general relativity nor on the
content of the universe or any assumption regarding cosmological
parameters. More accurate observation datasets and more effective
methods are still in need to make a clear answer on whether the
cosmic acceleration will keep increasing or not.

\end{abstract}


 \maketitle
 \renewcommand{\baselinestretch}{1.5}

\section{INTRODUCTION}\label{sec1}
More than ten years has passed since two independent groups
discovered that our universe entered a stage of accelerating
expansion using the Type Ia Supernova (SNIa) data~\cite{SNIa}.
During the past years, many observations embrace the results, such
as large scale structure~\cite{BAO}, the cosmic microwave background
(CMB) radiation~\cite{CMB}, and so on. All of these data strongly
suggest that we live in an accelerating universe, with the
deceleration parameter $q<0$ at low redshift, and an exotic
component named dark energy is introduced to interpret this
phenomena~\cite{DE}. As more and more observational data released,
the cosmic expansion history can be more accurately
determined~\cite{history, Daly}.

Besides the eternal acceleration scenario, such as the well-known
$\Lambda$CDM model, some authors have proposed that a matter
dominated decelerating expansion will resume soon after the
acceleration starts, because the vacuum energy's anti-gravitational
properties will reversed~\cite{Barrow}. Using the CPL
parametrization~\cite{CPL}, Shafieloo et al.~\cite{slowing} recently
found that the cosmic acceleration might be slowing down. This
question is interesting, since it challenges the $\Lambda$CDM model,
which fits most of the current data very well and predicts an
eternal cosmic acceleration. But this result is far from final
judgement. First, different parametrization methods of dark energy
equation of state will show different influences on the evolution
behavior of the universe~\cite{difference1}, and  the CPL ansatz
might be unable to fit the data simultaneously at low and high
redshifts. Second, some tensions also might exist in the SNIa
dataset which may lead to different results in the joint
analysis~\cite{Li, tension}, with one suggesting a decreasing
acceleration, others just the opposite. In order to avoid these
disadvantages, one can consider the so-called cosmographic
approach~\cite{cosmographic}, by directly parameterizing the
deceleration parameter, such as linear expansion
$q(z)=q_0+q_1z$~\cite{qz1}, or other two-parameter model
$q(z)=\frac{1}{2}+\frac{q_1z+q_2}{(1+z)^2}$~\cite{qz2}. And some
authors discovered that the universe might have already entered a
decelerating expansion era by using only low (for example $z<0.1$ or
$z<0.2$) redshift SNIa data~\cite{Wu2010, Guimaraes}.

In this paper, we focus on the deceleration parameter as the
diagnostic of the cosmological accelerating expansion. Following the
approach introduced in~\cite{Guimaraes}, we investigate the
deceleration parameter by expanding the luminosity distance to the
fourth order of redshift and the so-called
$y$-redshift~\cite{yredshift}, and fitting the latest SNIa sample
(Union2) released by the Supernova Cosmology Project (SCP)
Collaboration~\cite{Union2}, which contains 557 data including
recent large samples from other surveys and uses SALT2 for SNIa
light-curve fitting. But instead of cutting off the data at low
redshift, we use the sub-samples with $z<1.0$ throughout the paper,
in order to circumvent the convergence issues discussed
in~\cite{convergence} and to make our discussion physically
meaningful, while applying some other technique to reduce the
accumulation of systematic error, also we introduce some physical
constraint conditions to make our constraint more
robust~\cite{latest}. To make an overall comparison, the Hubble
parameter $H(z)$ is also expanded to the third order of the redshift
and the $y$-redshift and fitted to the Hubble evolution
data~\cite{Hubble, Hubble2}. This constitutes the second part of the
paper. In Sec. 3, we propose a non-parametrization method to
reconstruct the deceleration parameters at differen redshifts
directly from the SNIa data, and this method is independent of the
calibration of the SNIa data. Sec. 4 contains our conclusion.

\section{COSMOGRAPHIC EVALUATION}\label{sec2}
\subsection{Methodology}

We adopt the assumption that the universe has a flat
Friedmann-Robertson-Walker(FRW) metric and then the scale factor can
be approximated by the first several terms in a Taylor series; i.e.,
as a fifth order polynomial~\cite{expansion}. The coefficients in
this expansion are considered as the parameters of the theory. Since
one more term taken will greatly reduce the constraint ability of
the parameter (see Table 1 in the first paper of~\cite{xia}) and
four-parameter series can give a better approximation to the Hubble
parameter and the distance modulus than other
expansions~\cite{latest}. Therefore we take the first four terms
only in our analysis.

Accordingly, the luminosity distance $d_L(z)$ can be expanded to the
fourth order of redshift~\cite{cosmographic, Guimaraes, expansion},
which is a generalized form of the Hubble law,
\begin{eqnarray}
d_{L}(z)&=&\frac{1}{H_{0}}[z+\frac{1}{2}(1-q_{0})z^{2}-\frac{1}{6}(1-q_{0}-3q_{0}^{2}+j_{0})z^3\nn\\
&&+\frac{1}{24}(2-2q_0-15q_0^{2}-15q_0^{3}+10q_0j_0+5j_0+s_0)z^4]+\mathcal{O}(z^5)\,,
\label{eq:dlz}
\end{eqnarray}
where the coefficients are expressed with $H_0, q_0, j_0$, and
$s_0$,
 the present values of Hubble, deceleration, jerk and
snap parameters, respectively.  This kinematic method can gives
valuable information regarding the expansion rates of the universe,
thus provides us with a testing ground for all cosmological
solutions. To circumvent the convergence issues as discussed
in~\cite{convergence}, we choose the redshift cutoff $z<1.0$ with
537 SNIa events involved.

 A so-called $y$-redshift is introduced in
\cite{convergence}, where $y=\frac{z}{1+z}$. In this case, the
luminosity distance can be expanded as
\begin{eqnarray}
d_{L}(y)&=&\frac{1}{H_{0}}[y+\frac{1}{2}(3-q_{0})y^{2}+\frac{1}{6}(11-5q_{0}+3q_{0}^{2}-j_{0})y^3\nn\\
&&+\frac{1}{24}(50-26q_0+21q_0^{2}-15q_0^{3}+10q_0j_0-7j_0+s_0)y^4]+\mathcal{O}(y^5)\,.
\label{eq:dly}
\end{eqnarray}
This expansion converges faster than the expansion with the redshift
$z$ and the systematic error in this case is much smaller than the
former case. This expansion has been extensively studied
in~\cite{Guimaraes}.

In principle, the luminosity distance $d_L$ at low redshift $(z<1)$
can be exactly expressed by the infinite power series expansion of
$z$ or $y$, yet with infinite parameters involved. In practices, we
choose the first four terms of the Taylor series as an
approximation, which leads to the truncation error in our analysis.
This is considered as the systematic error. The systematic error in
(\ref{eq:dly}) is relatively larger compared with the Union2 data.
For example, when $y=1/2$, the systematic error in \ref{eq:dly} is
about $3.1\%$, while it is $1.0\%$ in the SNIa data. In order to
make the systematic error smaller than that of the Union2 data, one
has to cut off the data with $z\geq0.398$. Here we employ another
method to deal with this issue, rather than striking off high-$z$
SNIa data. First, we separate the data ($z<1.0$) equally in two
redshift bins and the divide is $z_{cut}=0.274$, to guarantee that
there are enough data in each bin. And then we employ
Eqs.~(\ref{eq:dlz}) and (\ref{eq:dly}) in each redshift bin,
respectively. Note that for the $z>z_{cut}$ case, the luminosity
distance can be expanded around $z=\frac{1}{2}$, which separates the
data ($0.274<z<1$) equally in two parts. This method will reduce the
systematic error and give a better constraint. Accordingly, the
luminosity distance for $z>z_{cut}$ can be expressed as
\begin{eqnarray}
d_L(z>z_{cut})&=&\frac{1}{H_0}[\frac{1}{24}(2-2q_0-15q_0^2-15q_0^3+10q_0j_0+5j_0+s_0)(z-\frac{1}{2})^4\nn\\
&&-\frac{1}{6}(1-q_0-3q_0^2+j_0-\frac{1}{2}(2-2q_0-15q_0^2-15q_0^3+10q_0j_0+5j_0+s_0))\nn\\
&&(z-\frac{1}{2})^3+(\frac{1}{16}(2-2q_0-15q_0^2-15q_0^3+10q_0j_0+5j_0+s_0)-\frac{1}{4}\nn\\
&&(1-q_0-3q_0^2+j_0)+\frac{1}{2}(1-q_0))(z-\frac{1}{2})^2+(1+\frac{1}{2}(1-q_0)\nn\\
&&-\frac{1}{8}(1-q_0-3q_0^2+j_0)+\frac{1}{48}(2-2q_0-15q_0^2-15q_0^3+10q_0j_0+5j_0+s_0))\nn\\
&&(z-\frac{1}{2})]+d_L(\frac{1}{2})\,.
\label{eq:dlz2}
\end{eqnarray}
Using the $y$-redshift, the divide is $y_{cut}=0.215$ and the
luminosity distance for the $y>y_{cut}$ case can be expanded around
the point $y=\frac{1}{3}$,
\begin{eqnarray}
d_L(y>y_{cut})&=&\frac{1}{H_0}[\frac{1}{24}(50-26q_0+21q_0^2-15q_0^3+10q_0j_0-7j_0+s_0)(y-\frac{1}{3})^4\nn\\
&&+\frac{1}{6}(\frac{1}{3}(50-26q_0+21q_0^2-15q_0^3+10q_0j_0-7j_0+s_0)\nn\\
&&+11-5q_0+3q_0^2-j_0)(y-\frac{1}{3})^3+\frac{1}{2}(\frac{1}{3}(11-5q_0+3q_0^2-j_0)\nn\\
&&+\frac{1}{18}(50-26q_0+21q_0^2-15q_0^3+10q_0j_0-7j_0+s_0)+3-q_0)(y-\frac{1}{3})^2\nn\\
&&+(1+\frac{1}{3}(3-q_0)+\frac{1}{18}(11-5q_0+3q_0^2-j_0)+\frac{1}{162}\nn\\
&&(50-26q_0+21q_0^2-15q_0^3+10q_0j_0-7j_0+s_0)(y-\frac{1}{3})]+d_L(\frac{1}{3})\,.
\label{eq:dly2}
\end{eqnarray}
The systematic error of Eq.~(\ref{eq:dly2}) is about 0.411\%, which
is comparable with the most accurately detected SNIa data, whose
error is 0.212\%.

The Hubble parameter $H(z)$ can also be expressed with $H_0, q_0,
j_0, s_0$ by expanding it to the third order. Since $q(z)$ is
related to the second order derivative of $d_L(z)$, and the first
order derivative of $H(z)$, this expansion is worthwhile to make a
full analysis. Using the redshift and the $y$-redshift, the hubble
parameter can be expanded as
\begin{equation}
E(z)\equiv\frac{H(z)}{H_0}=1+(1+q_0)z+\frac{1}{2}(j_0-q_0^2)z^2+\frac{1}{6}(3q_0^2+3q_0^3-3j_0-4q_0j_0-s_0)z^3\,,
\end{equation}
and
\begin{equation}
E(y)\equiv\frac{H(y)}{H_0}=1+(1+q_0)y+\frac{1}{2}(2+2q_0+j_0-q_0^2)z^2+\frac{1}{6}(6+6q_0-3q_0^2+3q_0^3+3j_0
-4q_0j_0-s_0)y^3\,.
\end{equation}
Similarly, in order to reduce the systematic error, we expand the
Hubble parameter in the regime of redshift $z>z_{cut}$ as
\begin{eqnarray}
\frac{H(z>z_{cut})}{H_0}&=&[1+q_0+\frac{1}{2}(j_0-q_0^2)+\frac{1}{8}(3q_0^2+3q_0^3-3j_0-4q_0j_0-s_0)](z-\frac{1}{2})\nn\\
&&+\frac{1}{2}[j_0-q_0^2+\frac{1}{2}(3q_0^2+3q_0^3-3j_0-4q_0j_0-s_0)](z-\frac{1}{2})^2\nn\\
&&+\frac{1}{6}(3q_0^2+3q_0^3-3j_0-4q_0j_0-s_0)(z-\frac{1}{2})^3+\frac{H(\frac{1}{2})}{H_0}\,,
\end{eqnarray}
and
\begin{eqnarray}
\frac{H(y>y_{cut})}{H_0}&=&[1+q_0+\frac{1}{3}(2+2q_0+j_0-q_0^2)+\frac{1}{18}(6+6q_0-3q_0^2+3q_0^3+3j_0-4q_0j_0\nn\\
&&-s_0)](y-\frac{1}{3})+\frac{1}{2}[2+2q_0+j_0-q_0^2+\frac{1}{3}(6+6q_0-3q_0^2+3q_0^3+3j_0-4q_0j_0\nn\\
&&-s_0)](y-\frac{1}{3})^2+\frac{1}{6}(6+6q_0-3q_0^2+3q_0^3+3j_0-4q_0j_0-s_0))(y-\frac{1}{3})^3\nn\\
&&+\frac{H(\frac{1}{3})}{H_0}\,.
\end{eqnarray}

\subsection{Datasets}

The dataset we use is the most recently released Union2 SNIa
dataset~\cite{Union2}, which contains 537 events for $z<1.0$. We
fit the SNIa by minimizing the $\chi^2$ value of the distance
modulus and the $\chi_{sn}^{2}$ for SnIa is obtained by comparing
theoretical distance modulus
$\mu_{th}(z)=5\log_{10}[d_L(z)]+\mu_{0}$, where $\mu_{0}=42.384-5\log_{10}h$, with observed $\mu_{ob}$ of
supernovae:

\[
\chi_{sn}^{2}=\sum_{i}^{537}\frac{[\mu_{th}(z_{i})-\mu_{ob}(z_{i})]^{2}}{\sigma^{2}(z_{i})}\]
To reduce the effect of $\mu_{0}$, we expand $\chi_{sn}^{2}$ with
respect to $\mu_{0}$ \cite{Nesseris:2005ur} :
\begin{equation}
\chi_{sn}^{2}=A+2B\mu_{0}+C\mu_{0}^{2}\label{eq:expand}\end{equation}
 where \begin{eqnarray*}
A & = & \sum_{i}\frac{[\mu_{th}(z_{i};\mu_{0}=0)-\mu_{ob}(z_{i})]^{2}}{\sigma^{2}(z_{i})},\\
B & = & \sum_{i}\frac{\mu_{th}(z_{i};\mu_{0}=0)-\mu_{ob}(z_{i})}{\sigma^{2}(z_{i})},\\
C & = & \sum_{i}\frac{1}{\sigma^{2}(z_{i})}.\end{eqnarray*}
 Eq.~(\ref{eq:expand}) has a minimum as \[
\widetilde{\chi}_{sn}^{2}=\chi_{sn,min}^{2}=A-B^{2}/C \]
 which is independent of $\mu_{0}$. In fact, it is equivalent to performing an
uniform marginalization over $\mu_{0}$, the difference between
$\widetilde{\chi}_{sn}^{2}$ and the marginalized $\chi_{sn}^{2}$ is
just a constant \cite{Nesseris:2005ur}. We will adopt
$\widetilde{\chi}_{sn}^{2}$ as the goodness of fit between
theoretical model and SNIa data.

In order to discriminate the evolution behaviors of different
models, another dataset is also employed, which is the recently
summarized Hubble evolution data~\cite{Hubble}, containing 7 data at
redshift $z<1.0$. We add in three more data at $z=0.24, 0.34, 0.43$
by taking the BAO scale as a standard ruler in the radial
direction~\cite{Hubble2}, and also $H(z=0)=74.2\pm3.6(km/(s\cdot
Mpc))$ observed by the Hubble Space Telescope (HST)~\cite{HST}. The
$\chi_{H}^{2}$ is defined as

\[
\chi_{H}^{2}=\sum_{i=1}^{11}\frac{[H(z_{i})-H_{ob}(z_{i})]^{2}}{\sigma_{i}^{2}}.
\]

To make use of the data, we perform a uniform marginalization over
the nuisance parameter $H_0$ and get

\[
\widetilde{\chi}_{H}^{2}=\chi_{H,min}^{2}=a-c^{2}/b, \]
  where \begin{eqnarray*}
a & = & \sum_{i}\frac{H_{ob}(z_{i})^{2}}{\sigma^{2}(z_{i})},\\
b & = & \sum_{i}\frac{[H_{th}(z_{i})/H_0]^{2}}{\sigma^{2}(z_{i})},\\
c & = &
\sum_{i}\frac{H_{ob}(z_{i})H_{th}(z_{i})/H_0}{\sigma^{2}(z_{i})}.
\end{eqnarray*}

\subsection{Results}

The analysis is performed by using the Monte Carlo Markov Chain in
the multidimensional parameter space to derive the likelihood. It is
natural to employ some physically obvious limitations to make the
estimation of parameters more robust~\cite{latest}. We impose the
positiveness conditions of the Hubble parameter and the luminosity
distance when we move randomly in the parameter space, and select
the points satisfying these conditions as our chains. Note that
these conditions are independent of the content of the universe and
the requirement of matter density fraction, In this sense they are
robust and the priors are:
\begin{eqnarray*}
d_{L}(z)>0,\\
H(z)>0.\end{eqnarray*}

\begin{center}
 \begin{figure}[htbp]
 \centering
\includegraphics[width=0.65\textwidth, height=0.45\textwidth]{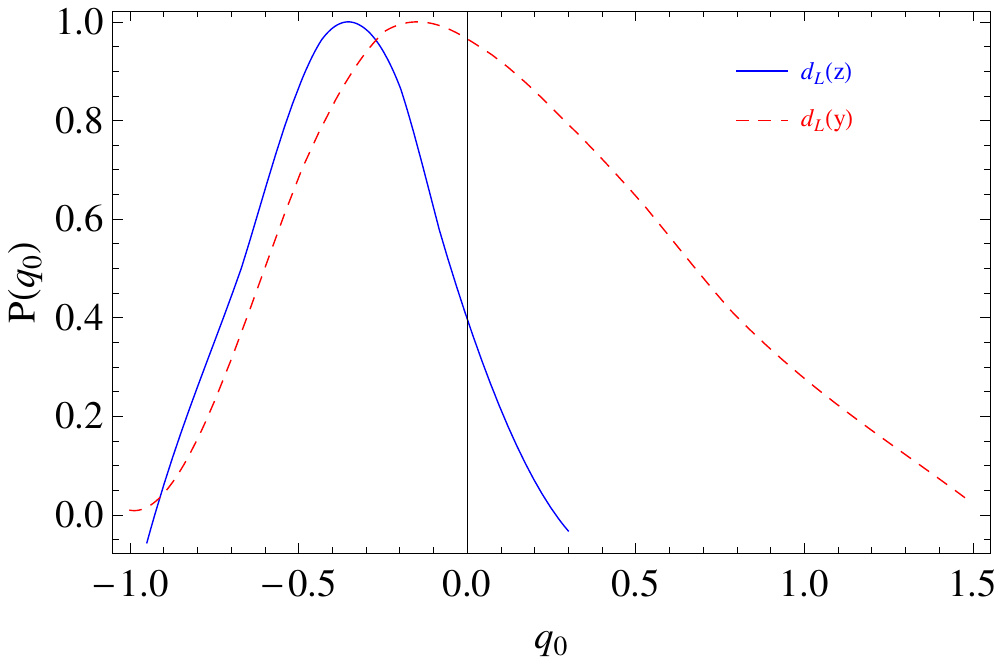}
\caption{\label{fig:likelihood}1D marginalized distribution
probability for the present deceleration parameter from the SNIa
data. Curves are scaled to $max[P(q_0)]=1.$}
\end{figure}
 \end{center}

\begin{center}
 \begin{figure}[htbp]
 \centering
\includegraphics[width=0.65\textwidth, height=0.45\textwidth]{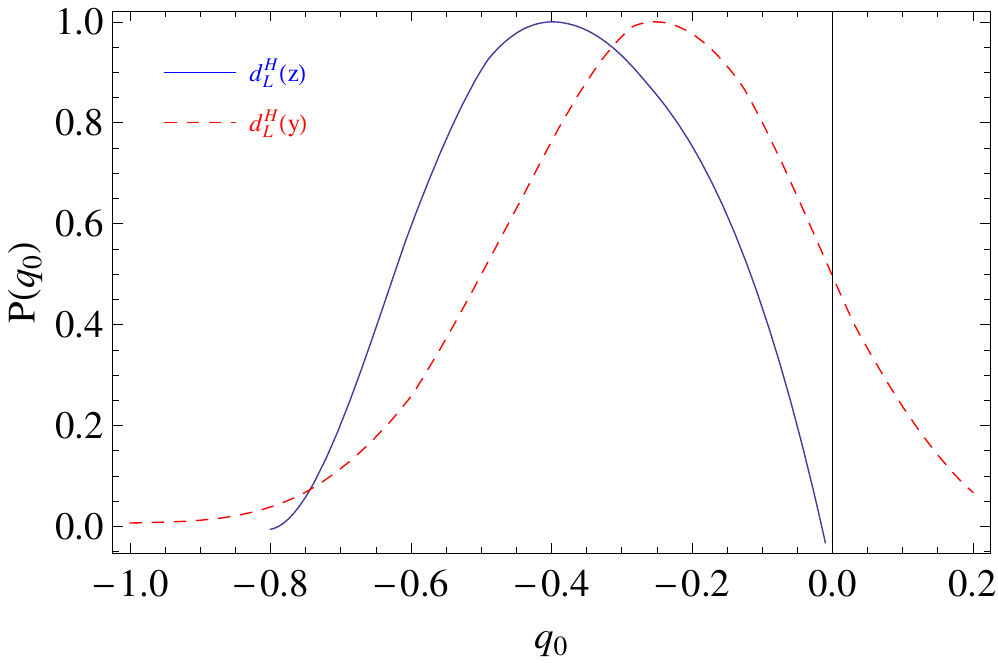}
 \caption{\label{fig:likelihoodh}1D marginalized distribution
probability for the present deceleration parameter from the SNIa
data and the Hubble evolution data. Curves are scaled to
$max[P(q_0)]=1.$}
\end{figure}
 \end{center}

We first investigate the constraint on the model parameters using
the SNIa data, and add the Hubble evolution data for a comparison,
then we analyze the evolution behavior of the deceleration
parameters. The best-fitting values and errors of parameters for
$d_L(z)$, $d_L(y)$ and $d_L^{(H)}(z)$, $d_L^{(H)}(y)$ (where the
upper index $H$ denote the case with Hubble evolution data added)
are summarized in Table \ref{tab:fit_result}, in company with the
best-fitting values of $\Lambda CDM$ parameters. For the $\Lambda
CDM$ model, we use the same dataset, and the best-fitting value of
current matter density fraction parameter $\Omega_{m0}=0.268$, thus,
$q_0=\frac{3}{2}\Omega_{m0}-1=-0.598$. The marginalized likelihood
distribution of $q_0$ is also shown in Figure \ref{fig:likelihood}
and \ref{fig:likelihoodh}, respectively.

If we use the SNIa data only, we find that the best-fitting values
of the deceleration parameter are both negative, which are larger
than that of the $\Lambda CDM$ model, and we can not exclude the
case of $q_0>0$ even within $1\sigma$ error. What's more, the
present value of the acceleration rate shows a slight tendency of
decrease for the case of $y$-redshift expansion, compared with the
case of redshift expansion. Note that even though the $y$-redshift
expansion is systematically more accurate than the redshift
expansion, but the figure of merit (FoM) of $q_0$, $j_0$ and $s_0$
is smaller, it is 1.10 for redshift expansion and 0.02 for
$y$-redhift expansion. Compared to the redshift expansion, the
$y$-redshift expansion gives a weaker constraint on the snap
parameter. Recently, Xia et al.~\cite{xia} investigated the fifth
order parameter $c_0$ using the similar cosmographic approach and
found that the constraint on it is much weaker.

With the Hubble evolution data added, the constraint ability is
stronger, and the FoM for redshift and $y$-redshift expansion are
3.07 and 0.52, respectively. It worths noting that by employing the
FoM to compare the constraint ability of the two methods, one often
assumes that the parameters' log-likelihood surface can be
quadratically approximated around the maximum, and the likelihood
for the data is Gaussian. So, it may significantly mis-estimate the
size of the errors if these assumptions do not hold. On the other
hand, the Bayesian Evidence, $E$, provides us with a good method to
compare different models~\cite{Jeffreys}, which is not dependent on
those assumptions. And the ratio of Evidences for two models
$B_{12}\equiv E(M_1)/E(M_2)$, also known as the Bayes factor,
provides a measure with which to discriminate the models.  We
calculate the Bayesian Evidence ratio of the redshift and
$y$-redshift expansion methods, and find that for the case of SNIa
dada used only, $\ln B=-0.83$, while for the case with Hubble data
added, $\ln B=-0.37$. It looks that the $y-$redshift expansion is a
bit better.  According to the Jeffreys grades, however, the result
means that these two expansion methods are comparable in fitting the
Union2 dataset.  We see from the results that the redshift expansion
approach excludes the case $q_0>0$ at about 95.4\% confidence level,
which tells us that the present universe is still in the stage of
accelerating expansion. The constraint on the snap parameter is much
tighter in the case of redshift expansion, and the constraint
ability is improved in some degree with our method compared with
some existing studies~\cite{cosmographic, xia}.

\begin{table}[!h]
\begin{tabular}{|c|c|c|c|c|}
\hline parameter  & $q_0$  & $j_0$  & $s_0$ &$\chi^2$\tabularnewline
\hline \hline $d_L(z)$ & $-0.357^{-0.533,\,-0.58}_{+0.557,\,+0.57}$
& $-1.826^{-4.792,\,-5.293}_{+5.848,\,+5.975}$  &
$-1.878^{-9.468,\,-13.905}_{+20.484,\,+21.703}$ &
$524.117$\tabularnewline \hline \hline $d_L(y)$ &
$-0.150_{+0.915,\,+1.456}^{-0.560,\,-0.795}$ &
$-6.564_{+11.119,\,+14.520}^{-21.400,\,-26.131}$  &
$-51.042_{+93.324,\,+125.622}^{-79.602,\,-90.412}$ &
$524.086$\tabularnewline \hline \hline $d_L^{(H)}(z)$ &
$-0.403_{+0.318,\,+0.352}^{-0.338,\,-0.431}$ &
$-1.230_{+3.218,\,+3.676}^{-3.468,\,-4.374}$  &
$-2.393_{+6.926,\,+7.891}^{-4.631,\,-9.281}$ &
$527.044$\tabularnewline \hline \hline $d_L^{(H)}(y)$ &
$-0.255_{+0.269,\,+0.460}^{-0.248,\,-0.445}$ &
$-7.631_{+5.110,\,+8.528}^{-6.185,\,-8.842}$ &
$-65.741_{+35.462,\,+72.343}^{-51.988,\,-80.440}$
 & $528.844$\tabularnewline \hline\hline $\Lambda CDM$ &
$-0.598$ & $1$  & $-0.206$
 & $525.765$\tabularnewline \hline
\end{tabular}
\tabcolsep 0pt \caption{\label{tab:fit_result}The best-fitting
values with $1\sigma$ and $2\sigma$ errors of $q_0, j_0$ and $ s_0$.
And the best-fitting values of $\Lambda CDM$ model fitted by the
same SNIa data.} \vspace*{5pt}
\end{table}


\begin{center}
 \begin{figure}[h]
 \centering
\includegraphics[width=0.8\textwidth, height=1\textwidth]{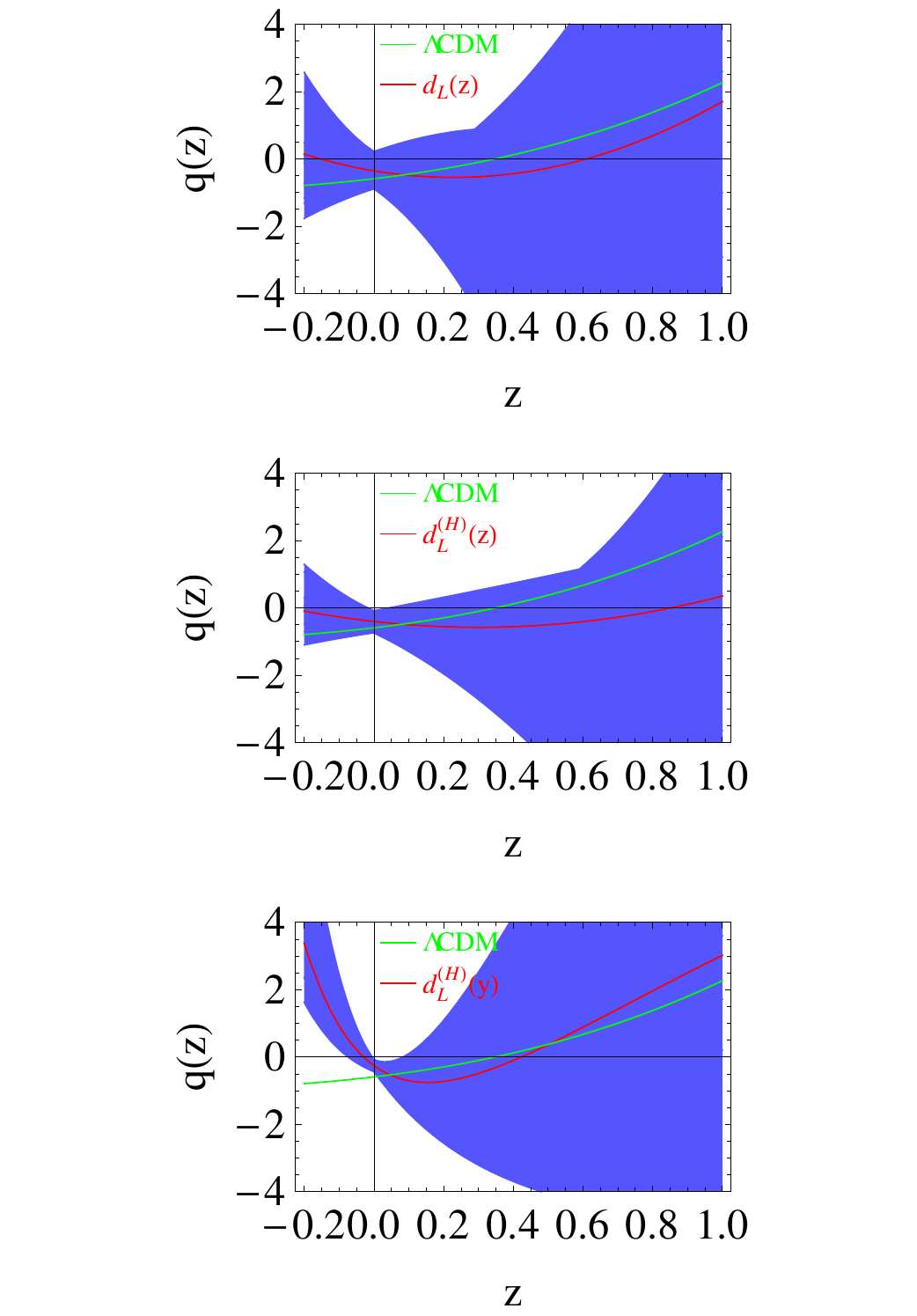}
\caption{\label{fig:q}Evolution behaviors of the deceleration
parameter. The red one is the best-fitting curve of $q(z)$ and the
green line is the evolution of $q(z)$ derived from the $\Lambda CDM$
model. The blue area delimits the 68.3\% confidence region for the
$q(z)$ reconstruction.}
\end{figure}
 \end{center}

It is of interest to investigate the evolution behavior of the
deceleration parameter to compare with the $\Lambda CDM$ model,
since different behaviors may tell something on the universe. And
another change in the sign of the cosmic acceleration is also
interesting to us, if any. Using the best-fitting parameters, the
deceleration parameter can be expanded with the redshift and
$y$-redshift, respectively, as
\begin{equation}
q(z)=q_0-(q_0+2q_0^2-j_0)z+\frac{1}{2}(2q_0+8q_0^2+8q_0^3-7q_0j_0-4j_0-s_0)z^2,
\end{equation}
and
\begin{equation}
q(y)=q_0-(q_0+2q_0^2-j_0)y+\frac{1}{2}(4q_0+8q_0^3-7q_0j_0-2j_0-s_0)y^2.
\end{equation}
We note that the quadratic form is essential for the detection of
the transient acceleration phase, since in this scenario, $q$ should
be positive in the past ($z>2$) and changes its sign at moderate
redshift ($z_t\sim0.3-1$), then steps into the phase of $q>0$. We
plot the evolution behaviors of $q$ compared with the $\Lambda CDM$
model in Figure \ref{fig:q}. Because of the weak constraint of the
$d_{L}(y)$, the $1\sigma$ area is too large to give any conclusive
information, therefore we do not show it here.

The best-fitting curves in the figures show a remarkably different
evolution behaviors of the deceleration parameter from the $\Lambda
CDM$ model, all of which indicate a transient acceleration. The
$\Lambda CDM$ model predicts that the universe began to accelerate
at $z=0.35$, but in our analysis, the redshift expansion predicts a
earlier transition and the $y$-redshift expansion predicts the
similar transition time as the $\Lambda CDM$ model. However, all
cases indicate a transient acceleration phase and predict a
decelerated stage in the near future $z<0$. This picture is even
clear in the case of the $y$-redshift expansion if both datasets are
used, which shows a considerably positive deceleration probability
today and in the future, at 68.3\% confidence level, and it
indicates that the $\Lambda CDM$ model is completely excluded in the
future. The behavior of transient acceleration is also predicted or
allowed by several dynamic models~\cite{slowing, model}. Due to
large uncertainties at high redshifts and in the future, however, it
is still not able to confirm the existence of transient acceleration
phase, nor the future decelerated acceleration, even at $1\sigma$
confidence level.

\begin{center}
 \begin{figure}[htbp]
 \centering
\includegraphics[width=0.65\textwidth, height=0.45\textwidth]{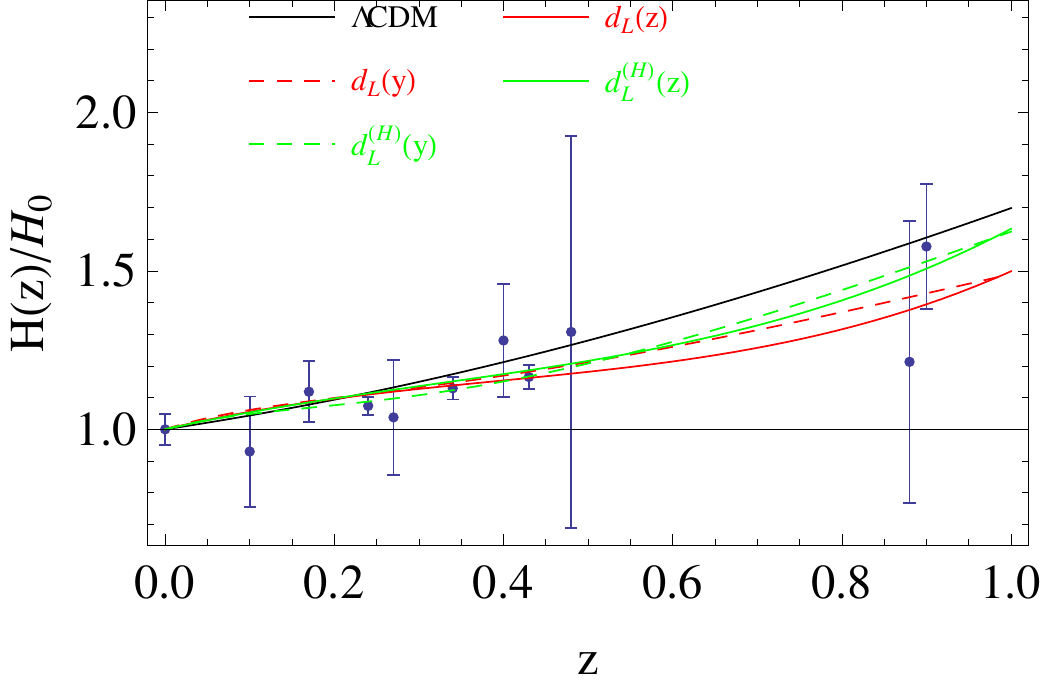}
 \caption{\label{fig:hubble}Best-fitting curves of $E(z)$ derived from the four cosmographic approaches. The red (or dashed red) line denotes the redshift (or
 $y$-redshift) expansion using SNIa data only. The green (or dashed green) line denotes the redshift (or
 $y$-redshift) expansion constrained with the Hubble data involved. The black one is the evolution of $E(z)$ derived from the
$\Lambda CDM$ model.}
\end{figure}
 \end{center}

Next, we plot the curves of $E(z)$ for the four cosmographic models
and the standard $\Lambda CDM$ model using the best-fitting
parameters, compared with the Hubble evolution data in Figure
\ref{fig:hubble}.  We see from the figure  that the curves are all
nicely consistent with the data, including the $\Lambda CDM$ model.
The two curves including the Hubble evolution data are much close to
the one of the $\Lambda CDM$ model than the other two. Note that 6
of the 11 Hubble evolution data have large error bars, with
$\sigma(obs)>10\%$, and the worst data is $\sigma(obs)=61.86\%$. So,
better measured data are in urgent need to make a definite
discrimination among different behaviors of our universe.

\section{Non-parametrization approach}\label{sec3}

Further exploration on the deceleration parameter is also approached
with a non-parametrization method, by reconstructing the
deceleration parameter from the distance modulus of SNIa directly,
which depends neither on the validity of general relativity nor the
content of the universe or any assumption regarding cosmological
parameters. From mathematics point of view, the determination of the
deceleration parameter is much difficult than the evolution of
Hubble parameter, since the Hubble parameter is the first order
derivative of the comoving distance, while the deceleration
parameter is related to the second order derivative of the comoving
distance. And because of that, the error for constructing the
deceleration parameter is also larger. To deal with this issue,
 Ref.~\cite{Daly} generated $2000$ artificial data sets mimicking what is
expected from SNIa measurement by the SNAP satellite
(see~\cite{SNAP} or the web site~\footnote{http://snap.lbl.gov}) and
then applied the quadratic expansion to express the dimensionless
coordinate distance $y(z)=H_0r(z)$ in each redshift bin, thus the
deceleration parameters can be obtained.

There are two sources of errors we try to tackle in the
implementation. First, the deviation from the true value of any
individual data is assumed to be normally distributed, which is a
good approximation only for relatively large samples. Second, any
finite dataset, especially small dataset, will contain large sample
variance. In summary, smaller number of data produces noisier
results, while larger number of data leads to more convictive
results at the expense of resolution in larger redshift bins. We try
to balance these effects by choosing relatively large number of SNIa
data in each redshift bin to make sure that there are both enough
data and not too large bins, and also to reduce the effect of sample
variance.

\begin{center}
 \begin{figure}[htbp]
 \centering
\includegraphics[width=0.65\textwidth, height=0.45\textwidth]{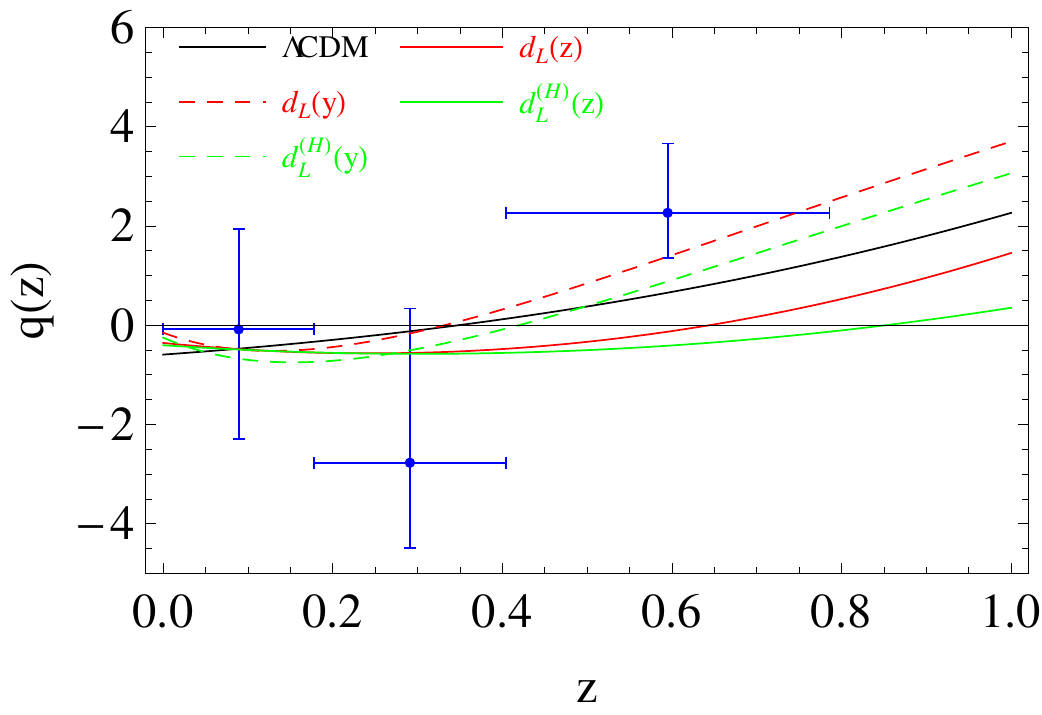}
\caption{\label{fig:qplot}Evolution of the deceleration parameter
reconstructed from the Union2 SNIa data with $1\sigma$ error bars.
Others are the best-fitting curves of the cosmographic approach as
before, in company with that of the $\Lambda CDM$ model.}
\end{figure}
 \end{center}

In our analysis, we focus on the Union2 dataset instead of the mock
datasets and separate the 557 SNIa data equally into four redshift
bins, each containing 139 data points (149 data points in the forth
bin). In each redshift bin, we reconstruct one data point using the
data within each bin, so we get totally 4 reconstructed data, as
shown in Table \ref{tab:mu}, where the first column is the averaged
redshift in each bin and the third column is the unbiased estimate
of the error. We will not take use of the expression of the distance
modulus of SNIa, which we believe contains some uncertainties in
determining the calibration of the data. Instead, we apply the
following expression,

\begin{equation}
\mu _{i+1}-\mu_{i}=5\lg\frac{d_L(z_{i+1})}{d_L(z_{i+1})},
\end{equation}
where $i$ runs from 1 to 3 in this case. This expression is credible
if the distance modulus is of logarithm form of luminous distance
only. In order to detect the deceleration parameter and make it
easier for numerical approach, we expand the comoving distance to
the second order of redshift in each redshift bin $z_i<z<z_{i+1}$,

\begin{equation}
r(z_{i+1})=r(z_i)+\frac{1}{H(z_i)}[z_{i+1}-z_i-\frac{1+q_i}{2(1+z_i)}(z_{i+1}-z_i)^2],
\end{equation}
where the Hubble parameter can also be expressed with the
deceleration parameter,
\begin{equation}
H(z_{i+1})=H(z_i)[1+\frac{1+q_i}{1+z_i}(z_{i+1}-z_i)].
\end{equation}
And then the luminosity distance can be obtained using the relation
$d_L(z)=r(z)(1+z)$,

\begin{equation}
d_L(z_{i+1})=\frac{1+z_{i+1}}{H(z_i)}[z_{i+1}-z_i-\frac{1+q_i}{2(1+z_i)}(z_{i+1}-z_i)^2]+
\frac{1+z_{i+1}}{1+z_i}d_L(z_i).
\end{equation}
For the beginning of the iteration process, we choose $q_0\approx
q_1$. In the following, we will see that this approximation is
reasonable since the redshift distance between them is small, with
$\delta z=0.089$. And we use the best-fitting value of Hubble
parameter given by the $WMAP7$ group, $h_0=0.704(km/(s\cdot
Mpc)$~\cite{CMB}. Then we get 3 deceleration parameters, shown in
Table \ref{tab:q}, where we choose the average redshift as the first
column. Since $q(z)$ involves the second order derivative of the
distance modulus, it is much harder to make the constraint
accurately, thus causes the error bars blowing up to be proportional
to $(\delta z)^{-5/2}$~\cite{error, Tegmark}, similar to the
equation of state of dark energy $w(z)$. We plot the evolution of
the reconstructed deceleration parameters in Figure \ref{fig:qplot}.
As a comparison, other four best-fitting curves of the cosmographic
approach are also shown. It is clear that except the $\Lambda CDM$
model, all other results contain a transient acceleration solution,
though the present value of the deceleration parameter is still
negative. We cannot discriminate their behaviors at low redshift
($z<0.4$), but higher redshift behaviors are distinct ($z>0.6$),
thus more accurately detected high-$z$ data are in urgent need. On
the other hand, we still need a well behaved method to affirm if
there is indeed a transient acceleration phase.
\begin{table}[!h]
\begin{tabular}{|c|c|c|}
\hline $z$  & $\mu$  & $\sigma_\mu$\\
\hline $0.10286$ & $35.3573$ & $0.1867$\\
\hline $0.17760$ & $39.4708$ & $0.1616$\\
\hline $0.40451$ & $41.7502$ & $0.2926$\\
\hline $0.78596$ & $43.4330$ & $0.3439$\\
\hline
\end{tabular}
\tabcolsep 0pt \caption{\label{tab:mu}Reconstructed distance modulus
of Union2 SNIa data, where the first column is the averaged redshift
in each bin and the third column is the unbiased estimate of the
error.} \vspace*{5pt}
\end{table}

\begin{table}[!h]
\begin{tabular}{|c|c|c|c|}
\hline $z$  & $q(z)$  & $\sigma_{+}$ & $\sigma_{-}$\\
\hline $0.089$  & $-0.0834$ & $+2.0285$ & $-2.2051$\\
\hline $0.291$ & $-2.7717$ & $+3.1068$ & $-1.7184$\\
\hline $0.595$ & $2.2651$ & $+1.3969$ & $-0.9106$\\
\hline
\end{tabular}
\tabcolsep 0pt \caption{\label{tab:q}Reconstructed deceleration
parameters of Union2 SNIa data, where the last two columns are the
estimates of the errors in each redshift bin, which are also
directly constructed from the SNIa data by considering the full
error of each data point.} \vspace*{5pt}
\end{table}

\section{CONCLUSION}\label{sec4}
To summarize, we have investigate the deceleration parameter in
detail, both with the cosmographic approach and the
non-parametrization method. These approaches depend neither on the
validity of general relativity nor the content of the universe. To
make the results robust, we focus on reducing the systematic error
throughout the paper.

Cosmographic approach to detect the deceleration parameter is to
expand the luminosity distance to the fourth order or even higher
order of redshift or the so called $y$-redshift, and then to fit the
datasets. But this method contains large uncertainties if we use the
high-$z$ data. Some authors cut off high-$z$ data in data fitting
process to avoid the problem, but we used other technique to reduce
the systematic error: By expanding the luminosity distance in two
redshift bins which contain the same number of SNIa data, the error
can be reduced. Since the deceleration parameter is related to the
second order derivative of the luminosity distance, and the first
order derivative of the Hubble parameter, it is also necessary to
expand the Hubble parameter to the third order as what we have done
for the luminosity distance, thus improving the constraint ability.
The result reveals that the universe may transit from decelerating
expansion to a transient accelerating expansion during $0.3<z<1.0$,
and the best-fitting evolution behaviors tell us that it is possible
that the universe will return to decelerating expansion stage in the
future even though it is still accelerating at present. Comparing
with the Hubble data shows that these results are indistinguishable
with that of the $\Lambda CDM$ model, which demands for more
accurate observations to come up with a conclusive verdict.

Non-parametrization method is also implemented by reconstructing the
deceleration parameters directly from the SNIa data. We simply use
the logarithmic form of the distance modulus with respect to the
luminosity distance, thus avoid the uncertainties in calibration of
the data. To reduce the systematic error, we distribute the same
number of SNIa data in each redshift bin to make sure of enough data
and not too large width of redshift bin. The result also indicates a
transient accelerating expansion stage of our universe. Since the
deceleration parameter is much more sensitive to the accuracy of the
SNIa data than the Hubble parameter, it is not a easy job to
constrain the result tightly, thus needs further improvement, both
in observation data and fitting methods.

\section*{ACKNOWLEDGEMENTS}
ZLT is really grateful to Bin Hu, Qiping Su and Hongbo Zhang for
their useful suggestions. RGC thanks the organizers and participants
for various discussions during the workshop ``Dark Energy and
Fundamental Theory" supported by the Special Fund for Theoretical
Physics from the National Natural Science Foundation of China with
grant No. 10947203. This work was supported in part by the National
Natural Science Foundation of China (No. 10821504, No. 10975168,
No.11035008 and No.11075098),  the Ministry of Science and
Technology of China under Grant No. 2010CB833004, and a grant from
the Chinese Academy of Sciences.

\end{document}